\documentclass[pre,aps,twocolumn,showpacs]{revtex4}
\usepackage{epsfig}
\usepackage{bm}

\begin{document}

\title{Broken chaotic clocks of brain neurons and depression}

\author{\small  A. Bershadskii}

\affiliation{\small {ICAR, P.O.B. 31155, Jerusalem 91000, Israel}}

\begin{abstract}
Irregular spiking time-series obtained {\it in vitro} and {\it in vivo} 
from singular brain neurons of different types of rats are analyzed by mapping to telegraph signals. 
Since the neural information is coded in the length of the interspike intervals 
and their positions on the time axis, this mapping is the most direct 
way to map a spike train into a signal which allows a proper application of the Fourier transform 
methods. This analysis shows that healthy neurons firing has periodic and chaotic deterministic 
clocks while for the rats representing genetic animal model of human depression these neuron 
clocks might be broken, that results in decoherence between the depressive neurons firing. 
Since depression is usually accompanied by a narrowing of consciousness this specific 
decoherence can be considered as a cause of the phenomenon of the consciousness narrowing as well. 
This suggestion is also supported by observation of the large-scale chaotic coherence of the posterior piriform 
and entorhinal cortices' electrical activity at transition from anesthesia to the waking state with 
full consciousness.           

\end{abstract}


\maketitle

\section{Introduction}

  All types of information, which is received 
by sensory system, are encoded by nerve cells 
into sequences of pulses of similar shape 
(spikes) before they are transmitted to the 
brain. Brain neurons use such 
sequences as main instrument for intercells 
connection. The information is reflected in 
the time intervals between successive firings 
(interspike intervals of the action potential 
train, see Fig. 1). There need be no loss of 
information in principle when converting from 
dynamical amplitude information to spike trains 
\cite{sau} and the irregular spike sequences 
are the foundation of neural information 
processing. Although understanding of the origin 
of interspike intervals irregularity has important 
implications for elucidating the temporal 
components of the neuronal code and for treatment 
of such mental disorders as depression and 
schizophrenia, the problem is still very far from 
its solution. 

The mighty Fourier transform method, for instance, is practically non-applicable to the 
spike time trains. The spikes are almost identical to each other and the neural information 
is coded in the length of the interspike intervals and the interspike intervals positions 
on the time axis, therefore it is the most direct way to map the spike train 
into a telegraph time signal, which has values -1 from one side of a spike and values +1 from 
another side of the spike with a chosen time-scale resolution. An example of such mapping is given in figure 1. 
While the information coding is here the same as for the corresponding spike train, the Fourier transform method 
is quite applicable to analysis of the telegraph time-series. 

On the other hand, recent dynamical models of neuron activity revealed new and complex role of regimes 
of a (deterministic) chaotic irregularity in the neuron spike trains (see, for instance, \cite{iz}-\cite{kf}). 
Therefore, we have to use all available mathematical tools in order to study the experimental data 
on the deterministic chaos properties (and, especially, in order to separate between deterministic chaos and 
stochasticity in the experimental signals).

\begin{figure} \vspace{-0.5cm}\centering
\epsfig{width=.45\textwidth,file=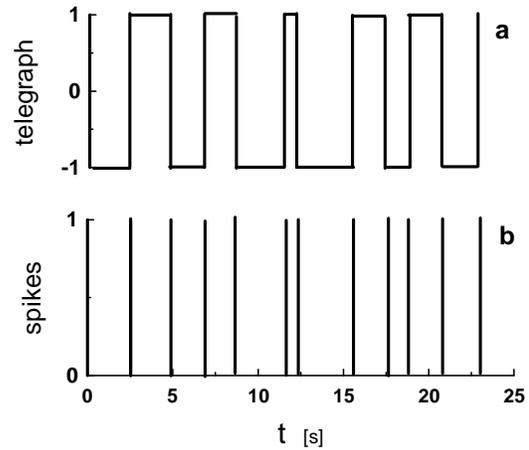} \vspace{-5cm}
\caption{Mapping of a spike train (figure 1b) into a telegraph signal (figure 1a).  }
\end{figure}

In present paper we have analyzed three types of experimentally obtained spike trains: 
a) obtained {\it in vitro} from a spontaneous activity in CA3 hippocampal slice culture of 
a healthy Wistar/ST rat (the raw data and the detail description of the experiment 
can be found online at http://hippocampus.jp/data and in Refs. \cite{sas},\cite{taka}), 
b) obtained in an electrophysiological {\it in vivo} experiment from neurons 
belonging to red nucleus of a healthy (Sprague-Dawley) rat's brain (see Ref. \cite{b1} 
for more details of the experiment and a preliminary discussion of the data), 
and c) obtained in an electrophysiological {\it in vivo} experiment from neurons belonging 
to red nucleus of a genetically depressed (Flinders Sensitive Rat Line) rat's brain (see Ref. \cite{b1} 
for more details and a preliminary discussion of the data). 

In the {\it in vitro} experiment a) a functional imaging technique with multicell loading of the calcium 
fluorophore was used in order to obtain the spike trains of spontaneously active singular neurons 
in the absence of external input \cite{sas},\cite{taka}. In the {\it in vivo} experiments 
b) and c) the rats were anesthetized and the extracellular 
recordings were processed from the singular cells \cite{b1}. \\

Motivation to study the hippocampus and red nucleus areas of brain in relation to the psychomotor 
aspects of depression is based on the recently discovered evidences of their deep involvement in this 
mental disorder. The hippocampus is a significant part of a brain system responsible for behavioral 
inhibition and attention, spatial memory, and navigation. It is also well known that spatial memory 
and navigation of the rats is closely related to the rhythms of their moving activity. 
On the other hand, the hippocampus of a human who has suffered long-term clinical depression can be 
as much as 20\% smaller than the hippocampus of someone who has never been depressed \cite{brem}. 
Inputs to the Red Nucleus arise from motor areas of the brain and in particular the deep cerebellar nuclei
(via superior cerebellar peduncle; crossed projection) and the motor cortex (corticorubral; ipsilateral projection).
On the other hand it is known that humans with deep depression have intrinsic locomotor's problems. 
Therefore, investigation of Red Nucleus for genetically defined rat model of depression (these rats partially
resemble depressed humans because they exhibit reduced appetite and psychomotor function) can be useful for
understanding the mental disorder origin.

\section{Neuron clock}
\begin{figure} \vspace{-1cm}\centering
\epsfig{width=.45\textwidth,file=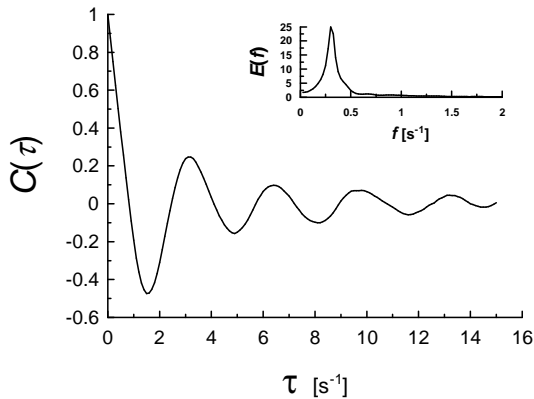} \vspace{-5cm}
\caption{Autocorrelation function for the telegraph signal corresponding to the cell-21 (800 spikes). Insert in the Fig. 2 shows corresponding spectrum. }
\end{figure}
\begin{figure} \vspace{-1cm}\centering
\epsfig{width=.45\textwidth,file=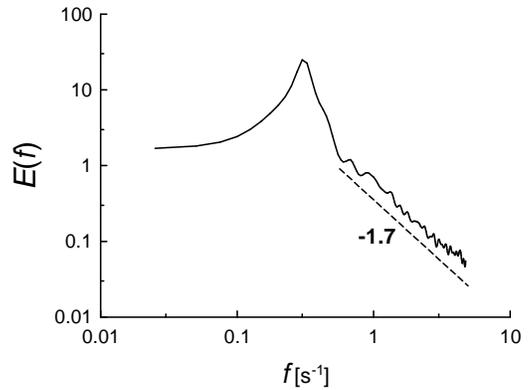} \vspace{-5cm}
\caption{Spectrum of the telegraph signal corresponding to the cell-21 in log-log scales. The dashed straight line 
indicates a power law: $E(f) \sim f^{-1.7}$.  }
\end{figure}
\begin{figure} \vspace{-0.5cm}\centering
\epsfig{width=.45\textwidth,file=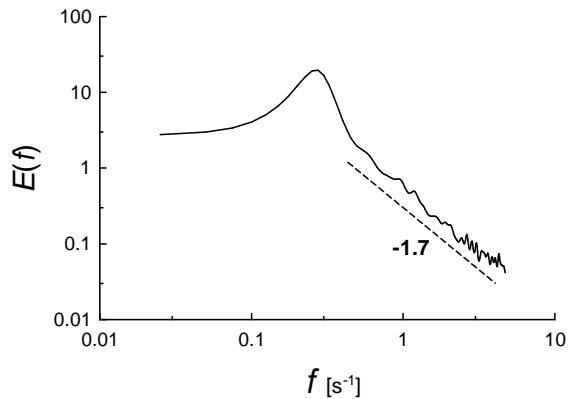} \vspace{-5cm}
\caption{As in Fig. 3 but for cell-25. }
\end{figure}

In the {\it in vitro} experiment with spontaneous activity of the hippocampal pyramidal cells 
different levels of activity were observed for different neurons \cite{sas},\cite{taka}. 
We take for our analysis the two most active neurons (http://hippocampus.jp/data - Data-006, 
cell-21, with 800 spikes in the time-series; and cell-25, with 692 spikes). The spike trains were mapped 
to telegraph signals as it is described above (see also \cite{bi}). Figure 2 shows autocorrelation function 
for the telegraph signal corresponding to the cell-21 (800 spikes). Insert in the Fig. 2 shows 
corresponding spectrum. Both the correlation function and the spectrum provide clear indication 
of a strong periodic component in the signal (the oscillations in the correlation function and 
the peak in the spectrum). The periodic component can be seen at frequency $f_0 \simeq 0.3$Hz. 
Figure 3 shows the spectrum in log-log scales. One can see that at high frequencies the spectrum 
exhibits a scaling behavior (power law: $E(f) \sim f^{-1.7}$, as indicated by the dashed straight line). 
The real power law can be more pronounced but under the experimental conditions individual spikes 
emitted at firing rates higher than 5Hz were experimentally inseparable \cite{sas},\cite{taka}. 
Figure 4 shows spectrum 
of the telegraph signal corresponding to the spike train obtained for the cell-25 (692 spikes). 
The spectrum is rather similar to the spectrum shown in Fig. 3 (for cell-21). The more broad peak in 
Fig. 4 can be related to the poorer statistics for the cell-25 in comparison with cell-21. The spectra 
were calculated using the maximum entropy method (because it
provides an optimal spectral resolution even for small data sets \cite{o}). 

\section{Threshold effect}

Now let us speculate about physics which could result in the spectra observed in Figs. 3 and 4. 
And let us recall some basic electrochemical properties 
of neuron. Nerve cells are surrounded by a membrane that 
allows some ions to pass through while it blocks the 
passage of other ions. When a neuron is not sending a 
signal it is said to be "at rest". At rest there are relatively more sodium ions onside the 
neuron and more potassium ions inside that neuron. The resting value of the {\it membrane} 
electrochemical potential $P$ (the voltage difference across the neural membrane) 
of a neuron is about -70mV. If some event 
(a stimulus) causes the resting potential to move toward 0mV and the depolarization reaches about -55mV 
(a "normal" threshold) a neuron will fire an {\it action} potential. 
The action potential is an explosive release of charge between neuron and its surroundings that is 
created by a depolarizing current. If the neuron does not reach this critical threshold level, 
then no action potential will fire. Also, when the threshold level is 
reached, an action potential of a {\it fixed} size will always fire (for any given neuron the size of 
the action potential is always the same). Depending on different types of 
voltage-dependent ion channels, different types of action 
potentials are generated in different cells types and the 
qualitative estimates of the potentials and time periods can 
be varied.
\begin{figure} \vspace{-0.5cm}\centering
\epsfig{width=.45\textwidth,file=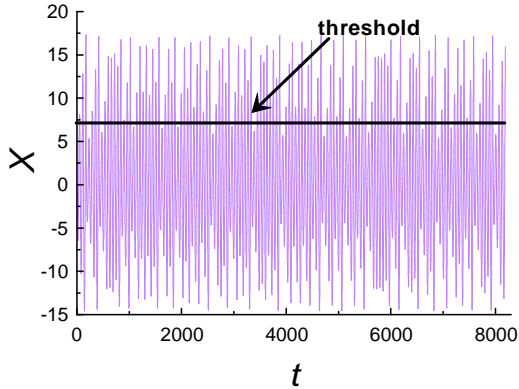} \vspace{-5.5cm}
\caption{X-component fluctuations 
of a chaotic solution of the R\"{o}ssler system Eq. (1) ($a=0.15,~ b=0.20,~ c=10.0$). }
\end{figure}
\begin{figure} \vspace{-0.3cm}\centering
\epsfig{width=.45\textwidth,file=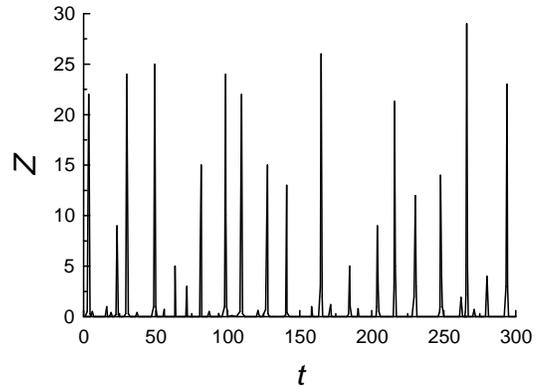} \vspace{-5.7cm}
\caption{Z-component fluctuations 
of a chaotic solution of the R\"{o}ssler system. }
\end{figure}
\begin{figure} \vspace{-0.5cm}\centering
\epsfig{width=.45\textwidth,file=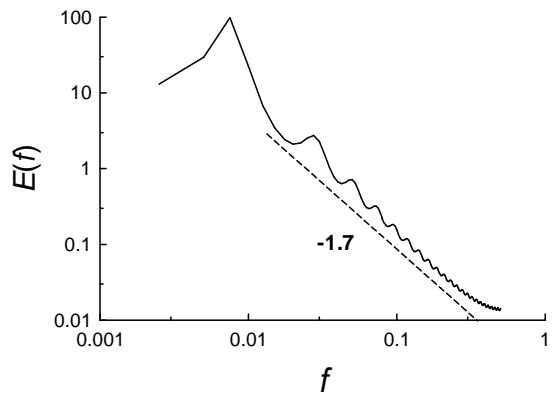} \vspace{-5cm}
\caption{Spectrum of the telegraph signal corresponding to the spike train generated by 
the x-component fluctuations overcoming the threshold $x=7$. 
The dashed straight line indicates a power law: $E(f) \sim f^{-1.7}$. }
\end{figure}
Recent reconstructions of a driver of the {\it membrane} potential using the 
neuron spike trains indicate the R\"{o}ssler oscillator as the most probable (and simple) 
candidate (see, for instance, Refs. \cite{bi},\cite{cas}-\cite{per}). 
Figure 5 shows as example the x-component fluctuations 
of a {\it chaotic} solution of the R\"{o}ssler system \cite{ross} 

$$
\frac{dx}{dt} = -y -z;~~  \frac{dy}{dt} = x + a y;~~  \frac{dz}{dt} = b + x z - c z \eqno{(1)}
$$      
where a, b and c are parameters). At certain values of the parameters a,b and c the $z$-component 
of the R\"{o}ssler system is a {\it spiky} time series \cite{llg},\cite{lgl}: Fig. 6. 
It can be shown that the R\"{o}ssler system and the 
well known Hindmarsh-Rose model \cite{hr} of neurons are subsystems of the same differential model with 
a spiky component \cite{lgl}. Previously the 'spiky' component of such models was interpreted and studied 
as a simulation of a neuronal {\it output}. For the {\it spontaneous} neuron firing 
(without external stimulus), however, we suggest to reverse the approach and consider the spiky variable 
as the main component of the electrical {\it input} (which naturally should have 
a 'spiky' character, see above) to the neuron under consideration. For each neuron the height of the 
spikes, which the neuron generates, is about the same. However, the heights of the spikes generated 
by different neurons are different. Also the signals coming from different neurons to the neuron under 
consideration have to go through the electrochemical passes with different properties. 
Therefore, the spiky $z$-time-series (Fig. 6) can naturally represent 
a multineuron signal, which can be considered as a spontaneous input for the neuron under consideration. 
If we use the usual interpretation 
of the $x$-component as a driver of the membrane potential $P(x)$ and the $y$-component as that taking into 
account the transport of ions across the membrane through the ion channels \cite{hr}, then the 
position of the input (the component $z$) in the first equation of the system Eq. (1) has a 
good physical background (cf. Ref. \cite{hr}). Then, the quadratic 
nonlinearity in the third equation of the system Eq. (1) can be interpreted as a simple
(in the Taylor expansion terms) feedback of the neuron to the main component of the neuronal input. 
This model with the strong nonlinear feedback can 
be relevant to the most active neurons of a spontaneously active brain (see below results of 
an {\it in vitro} experiment with a spontaneous brain activity). The details of the function $P(x)$ is not 
significant for the threshold firing process, what really matters is that the membrane potential 
function $P(x)$ reaches its firing value when (and only when) its argument $x$ crosses certain threshold 
from below. In this simple model the driving variable $x$ may overcome its threshold value (Fig. 5) 
due to the deterministic (chaotic) spontaneous stimulus. Let us consider an output spike signal resulting from 
overcoming a threshold value $x=7$, for instance. Fig. 7
shows spectrum of the telegraph signal corresponding to the spike signal. 

One can compare Fig. 7 with Figs. 3 and 4 to see very good reproduction of the main spectral 
properties.

In order to understand what is going on here we show in figure 8 spectrum of the x-component itself. 
\begin{figure} \vspace{-0.5cm}\centering
\epsfig{width=.45\textwidth,file=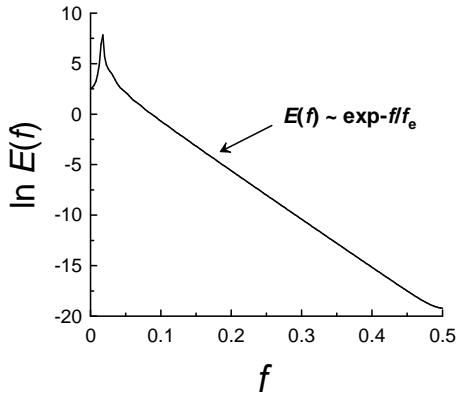} \vspace{-5cm}
\caption{Spectrum of the x-component fluctuations shown in Fig. 5. We used the semi-log 
axes in order to indicate exponential decay of the spectrum.
 }
\end{figure}

The semi-log scales are used in these figures in order to indicate exponential decay 
in the spectra (in the semi-log scales this decay corresponds to a straight line):
$$
E(f) \sim e^{-f/f_e}        \eqno{(2)}
$$
While the peak in the spectrum corresponds to the fundamental frequency, $f_0$, of the 
R\"{o}ssler chaotic attractor, the rate of the exponentional decay (the slope of the 
straight line in Fig. 8 provides us with and additional characteristic frequency $f_e$. 
Thus R\"{o}ssler chaotic attractor has two clocks: periodic with frequency $f_0$ and 
chaotic with frequency $f_e$. If one compares Fig. 8 and Fig. 7 one can see that 
the periodic clock survived the threshold crossing (with a period doubling, see Appendix). 
The chaotic clock, however, did not survive the threshold crossing at spontaneous activity: 
the exponential decay in Fig. 8 has been transformed into a scaling (power law) decay in Fig. 7, 
which has no characteristic frequency (scale invariance). The scaling exponent value '-1.7' is 
not sensitive to a reasonable variation of the threshold value ($\sim 20$\%) and even 
to Gaussian fluctuations of the threshold value. Therefore, it is not just a coincidence 
that the scaling law in the R\"{o}ssler case agrees with results of the {\it in vitro} experiment 
(cf. also \cite{all},\cite{luk},\cite{grig} and Fig. 16a).
\begin{figure} \vspace{-0.5cm}\centering
\epsfig{width=.45\textwidth,file=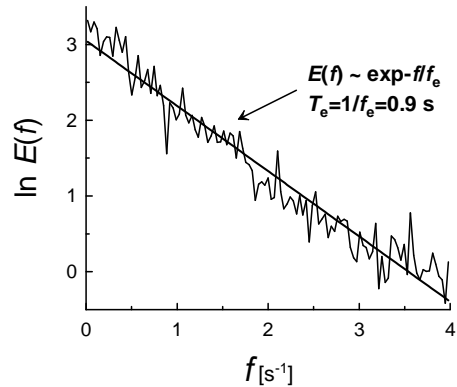} \vspace{-5cm}
\caption{Spectrum of the telegraph signal corresponding to a healthy red nucleus cell. The data is shown 
in the semi-log scales in order to indicate the exponential decay Eq. 2 (the straight line). 
}
\end{figure}
\begin{figure} \vspace{-0.5cm}\centering
\epsfig{width=.45\textwidth,file=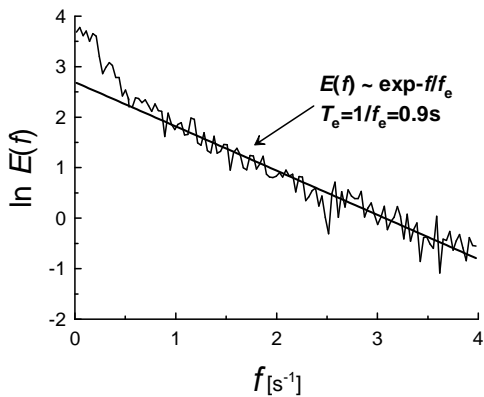} \vspace{-5cm}
\caption{As in Fig. 9 but for another healthy cell. }
\end{figure}
It should be noted that for a wide class of deterministic systems a broad-band spectrum 
with {\it exponential} decay is a generic feature of their {\it chaotic} solutions 
Refs. \cite{o},\cite{sig}-\cite{fm}. It is shown in Ref. \cite{sig} 
that the characteristic frequency 
$$
f_e = \sum_i \lambda^{+}_i  \eqno{(3)}
$$
where $ \lambda^{+}_i$ are positive Lyapunov exponents of the chaotic system.

\section{Chaos vs. stochasticity in neuron firing}

Both stochastic and deterministic processes can result in the
broad-band part of the spectrum, but the decay in the
spectral power is different for the two cases. An exponential decay with respect 
to frequency refers to chaotic time series while a power-law decay indicates that the spectrum 
is stochastic. 

Figure 9 shows a power spectrum obtained by the fast Fourier transform method 
applied to a telegraph signal mapped from a spike train measured in 
the red nucleus of a healthy rat (we can use the fast Fourier transform here due to sufficiently 
large number of spikes in the spike train: 2170). The spike train corresponds to a singular neuron firing. 
Figure 10 shows analogous spectrum obtained from another healthy red nucleus's neuron (2139 spikes). 
The semi-log scales are used in these figures in order to indicate exponential decay 
in the spectra (unlike the situation discribed above for a spontaneous activity): 
in the semi-log scales this decay corresponds to a straight line - Eq. 2. The characteristic 
frequency $f_e \simeq 1.1$Hz in the both cases.
\begin{figure} \vspace{-0.5cm}\centering
\epsfig{width=.45\textwidth,file=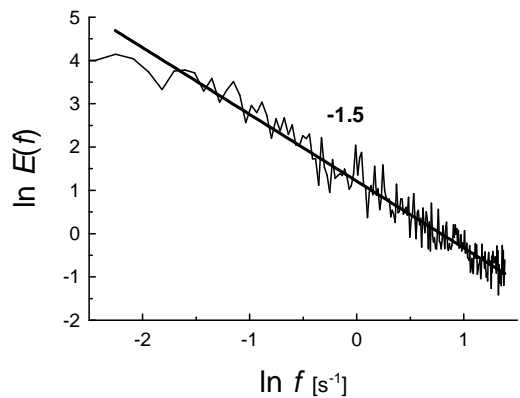} \vspace{-5cm}
\caption{Spectrum of the telegraph signal corresponding to a genetically depressed red nucleus cell. 
The data are shown in the log-log scales in order to indicate the power law decay Eq. 4 (the straight line). 
 }
\end{figure}
\begin{figure} \vspace{-0.5cm}\centering
\epsfig{width=.45\textwidth,file=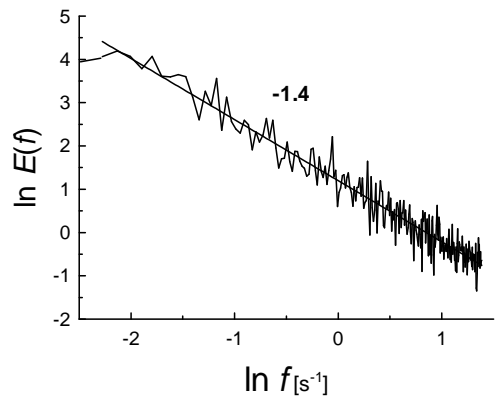} \vspace{-5cm}
\caption{As in Fig. 11 but for another genetically depressed red nucleus cell. }
\end{figure}
Figure 11 shows a power spectrum obtained by the fast Fourier transform method 
applied to a telegraph signal mapped from a spike train (2022 spikes) measured in 
the red nucleus of a genetically depressive (the "Flinders" line) rat. 
The spike train corresponds to a singular neuron firing. Figure 12 shows analogous spectrum 
obtained from another genetically depressive red nucleus's neuron (2048 spikes). 
The log-log scales are used in these figures in order to indicate a power-law decay 
in the spectra (in the log-log scales this decay corresponds to a straight line):
$$
E(f) \sim f^{-\alpha}       \eqno{(4)}
$$ 

In this (scaling) situation there is no characteristic time scale. 
The scaling exponent $\alpha \simeq 1.5\pm 0.1$ and $\simeq 1.4 \pm 0.1$ for these two cases.

\section{Chaotic neural coherence and depression}

In order to work together the brain neurons have to make adjustment of their rhythms. 
The main problem for this adjustment is the very noisy environment of the brain neurons. 
If their work was based on pure periodic inner clocks this adjustment would be 
impossible due to the noise. The nature, however, has another option. This option is 
a chaotic clock. In chaotic attractors certain characteristic frequencies can be embedded by 
broad-band spectra, that makes them much more stable to the noise perturbations \cite{ra}. 

In the light of presented results one can conclude 
that for the considered cases the healthy neurons firing 
has deterministic clocks (periodic and chaotic), while the genetically depressive red nucleus's 
neurons exhibited a pure stochastic firing and it seems that their background deterministic clocks were broken. 
The existence of the background clocks can be utilized by the healthy neurons for 
synchronization of their activity \cite{iz},\cite{sas},\cite{taka},\cite{aba}-\cite{ros}. 

In order to compare coherent properties of the healthy and the depressive neuron pairs we will use 
cross-spectral analysis. The cross spectrum $E_{1,2}(f)$ of two processes $x_1(t)$ and $x_2(t)$ is defined 
by the Fourier transformation of the cross-correlation function normalized by the product of square root of the univariate power spectra $E_1(f)$ and $E_2(f)$:
$$
E_{1,2}(f)= \frac{\sum_{\tau} \langle x_1(t)x_2(t-\tau) \rangle 
\exp (-i2\pi f \tau)}{2\pi \sqrt{E_1(f)E_2(f)}} \eqno{(5)}
$$
the bracket $\langle... \rangle$ denotes the expectation value. The cross spectrum can be decomposed
into the phase spectrum $\phi_{1,2} (f)$ and the coherency $C_{1,2}(f)$:
$$
E_{1,2}(f)= C_{1,2}(f) e^{-i \phi_{1,2} (f)}  \eqno{(6)}
$$
Because of the normalization of the cross spectrum the coherency
is ranging from $C_{1,2}(f)=0$, i.e. no linear relationship between
$x_1(t)$ and $x_2(t)$ at $f$, to $C_{1,2}(f)=1$, i.e. perfect linear relationship. 
\begin{figure} \vspace{-1cm}\centering
\epsfig{width=.45\textwidth,file=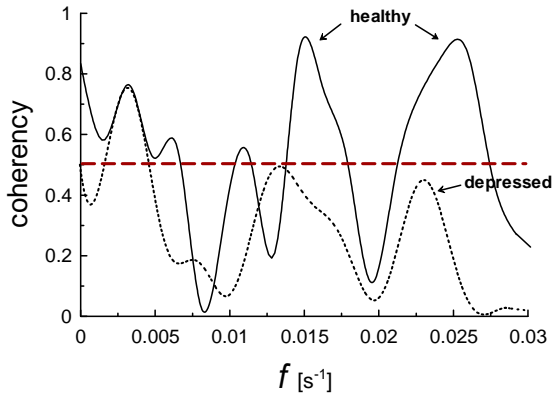} \vspace{-5cm}
\caption{Comparison of coherency in firing for the healthy (solid curve) 
and for the genetically depressed (doted curve) 
neuron pairs in a low-frequency domain (the {\it in vivo} experiments). }
\end{figure}
 
Figure 13 shows comparison of coherency in firing for the healthy (solid curve) and for the genetically depressed 
(doted curve) neuron pairs in a low-frequency domain (the {\it in vivo} experiments). 
Despite of the deep anesthesia the healthy 
neurons exhibit bands of rather high ($> 0.5$) coherency in the low-frequency domain, 
while the depressive neurons activity is rather decoherent in this domain.

\section{Long-range chaotic coherence}

\begin{figure} \vspace{-0.5cm}\centering
\epsfig{width=0.45\textwidth,file=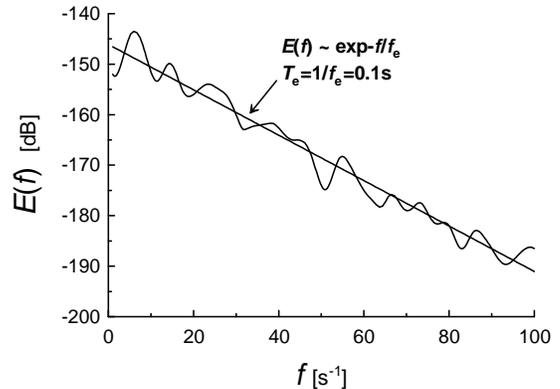} \vspace{-5.5cm}
\caption{Spectrum of local field potentials for the posterior piriform (the data were taken from Ref. \cite{hermer}). 
The data are shown in the semi-log scales in order to indicate the exponential decay Eq. (2) 
(the straight line).}
\end{figure}

The chaotic coherence can involve a large number of the healthy neurons and may be the entire brain. 
The multi-second oscillations, for instance, are known to be synchronized nearly brain-wide \cite{cont},\cite{ster}.
In the case of depression, however, the chaotic neuron clocks can be broken in a significant part of the brain 
neurons. That can result in certain decoherence in different parts of the brain. Since depression is usually 
accompanied by a narrowing of consciousness (and a distorted sense of time) this specific decoherence could be considered as a cause of the phenomenon of the consciousness narrowing as well. The coherence is important for attention, sensorimotor processing, etc.. In humans, in particular, being low in attentional flexibility 
magnified the effects of private self-focused attention so typical for depressive persons. \\

In order to support the possibility of the extended chaotic coherence we will use analysis of 
simultaneously recorded local field potentials from three sites along the olfactory-entorhinal axis (the anterior piriform, posterior piriform, and entorhinal cortices: aPIR, pPIR and Ent C) reported in a recent paper \cite{hermer}. The measurements reported in the Ref. \cite{hermer} 
were performed in lightly anesthetized healthy rats (the Long-Evans rats with electrode bundles implanted in their 
anterior and posterior cortices, and with vertical, silicon probes in their entorhinal cortices), which were emerged from the anesthesia to the waking state with full consciousness. Since the measured local field potentials time series are not spiky ones one does not need in the special data mappings in this case. The authors of the Ref. \cite{hermer} discovered a new form of coherent neural activity across the three widely separated brain sites, which they named Synchronous Frequency Bursts (SFBs). 
The high-energy bursts of spontaneous momentary synchrony were observed across widely separated olfactory and 
entorhinal sites (which have also different architecture: the 6 layers of the entorhinal cortex vs. the three layers of the piriform cortices). Moreover, a significant rate of the SFBs simultaneous occurrences was also observed across the different functional processing systems: motor and olfactory ones.\\

The stereotypical duration of the SFBs was about 250 ms and the power spectra taken across the events were exponentially decaying. Figure 14 shows a typical spectrum for the posterior piriform 
area. The straight line is drawn in this figure in order to indicate the exponential decay Eq. (2) in the semi-log scales (cf. Figs. 9 and 10 for the singular neuron firing). The exponential decay indicates a chaotic nature of these 
bursts (see above). The decay rate $T_e=1/f_e \simeq 0.1s$ is significantly smaller than that observed for the 
singular neurons (Figs. 9 and 10). Taking into account Eq. (3) one can conclude that the chaotic mixing in the phase space (determined by the Lyapunov's exponents) is much more active for the multineuron activity than for the 
singular neuron firing (that seems quite natural). This more active mixing shifts the spectrum into more high 
frequency range. Moreover, one can expect that expansion (globalization) of the chaotic coherence on the 
larger brain areas will shift the coherent chaotic activity even into the higher frequency ranges (cf. below). \\
\begin{figure} \vspace{-1cm}\centering
\epsfig{width=.45\textwidth,file=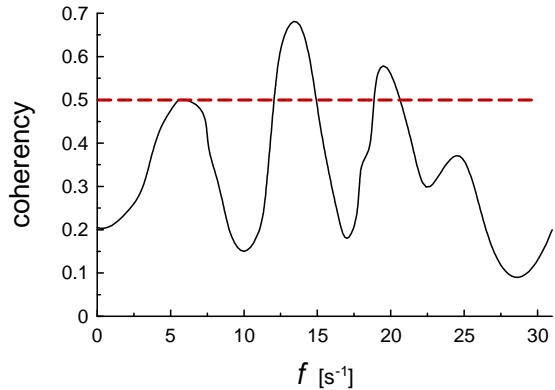} \vspace{-5cm}
\caption{Coherency calculated for the SFBs in the posterior piriform and entorhinal cortices 
(the data were taken from Ref. \cite{hermer}). }
\end{figure}
The authors of the Ref. \cite{hermer} computed coherence across SFBs in a pair of brain regions. Figure 15 
shows the coherency calculated for the SFBs in the posterior piriform and entorhinal cortices (which are
separated in brain space by about 8mm). One can compare 
this figure with the Fig. 13 (where the coherency was calculated for a pair of neighboring neurons). 
In this case the 
frequency bands of high coherency can be observed as well. The coherent frequency-range is shifted 
considerably in the high frequency direction for the multineuron case (see above for a reason of this shift). 
Actually, "the main purpose of SFBs might be to coordinate multiple frequency bands across different processing subsystems" \cite{hermer}. Such {\it coordination} provides a sufficient level of coherence for the work of these 
separated subsystems with speed and efficiency impossible in the case of transmission of a specific behavioral 
content. This can be considered as the main advantage of the chaotic coherence. The hardware for these effective 'management' can be provided (at least partially) by recently discovered in the cortex and hippocampus 
interneuronal networks with long-range axonal connections \cite{klaus},\cite{jin} and for the high frequency $\gamma$-range (30-90Hz) oscillations "via neurons (and glia) inter-connected by electrical synapses called gap junctions which physically fuse and electrically couple neighboring cells."\cite{ham2}.   \\

The authors of the Ref. \cite{hermer} observed also that the SFBs occurrence is a function of level of 
consciousness. They found "that the SFBs occurred far more often under light anesthesia than deeper anesthetic 
states, and were especially prevalent as the animals regained consciousness". 
They did not observe the SFBs after the rats regained full alertness, but as they comment this can be a technical problem of inferring the specific signal from the highly complex local field potential of the awake state. 
Therefore, one cannot rule out the possibility that the phenomenon is still in a full swing also in the fully consciousness state (at least at certain conditions).   \\

Finally, it should be noted that the transitional states of consciousness (emerging and decaying) have a very interesting relationship to associative human creativity (H. Poincare called these states as semi-somnolent 
conditions, see Ref. \cite{puan}, Chapter: Mathematical discovery). 
The very creative and unexpected {\it associative} ideas that come in these states can have the above described long-range chaotic coherence as their direct physical background. Moreover, the same mechanism can also be in work at full consciousness (see previous paragraph). In this case, however, its results are considered as ones coming from 
the 'clear sky' and we tend to interpret them (may be wrongly) as a result of a prolonged period of 
unconscious work. In the full consciousness state these results are more often turn out to be adequate ones, 
unlike of those obtained in the transitional states \cite{puan}).\\

"A new result has value, if any, when, by establishing connections between elements that are known but until then dispersed and apparently unrelated to one another, order is immediately created where chaos seemed to reign" \cite{puan}.

\section{ACKNOWLEDGMENTS}

I thank Dremencov E. and Ikegaya Y. 
for sharing the data and discussions and also Allegrini P. and Grigolini P. for comments and suggestions. 
I thank Greenberg A. for help in computing.

\section{Appendix}

In order to understand how the fundamental chaotic clock survives the threshold firing 
let us consider a very simple and rough model, which allows analytic calculation of its 
autocorrelation function. In this model the spike firing takes place at
discrete moments: $t_n=nT+\zeta$, where $\zeta$ is an uniformly distributed over the interval 
$[0, T]$ random variable, 
$n=1,2,3...$ and $T$ is a fixed period. Then let us consider a 
telegraph signal constructed for this spike train as it has been described above. If $p$ 
is a probability of the spike firing at a current moment ($0 \leq p < 1$), then the autocorrelation 
function of such telegraph signal is:
$$
C(\tau) = (n-\tau/T)(2p-1)^{n-1}+(\tau/T-(n-1))(2p-1)^n  \eqno{(A1)}
$$
in the interval $(n-1)T \leq \tau < nT$. Figure 16b shows the autocorrelation function Eq. (A1) calculated 
for $p=0.25$, as an example. For comparison figure 16a shows also a superposition of the autocorrelation functions for the telegraph signals corresponding to the cell-21 (the solid line) and to the spike train generated by 
the the R\"{o}ssler attractor fluctuations overcoming the threshold $x=7$ (circles). In order to make 
the autocorrelation functions comparable a rescaling has been made for the R\"{o}ssler attractor generated autocorrelation function.
\begin{figure} \vspace{-1cm}\centering
\epsfig{width=.45\textwidth,file=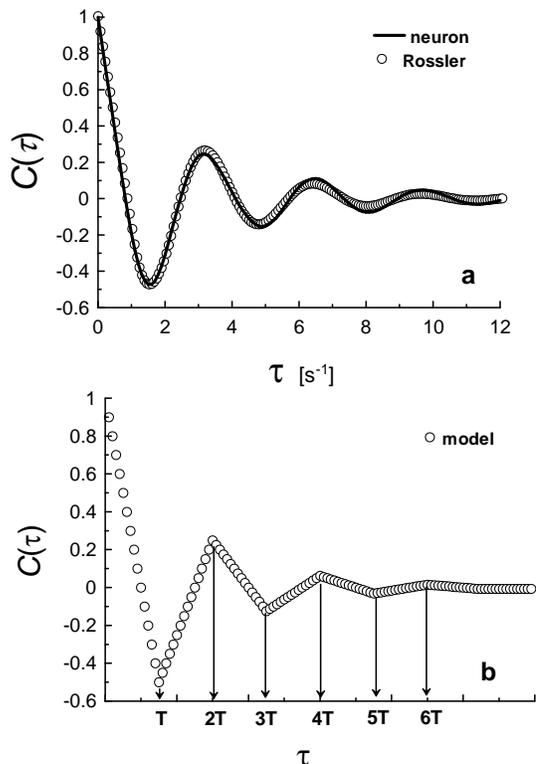} \vspace{-1cm}
\caption{
{\bf a}) Autocorrelation functions for the telegraph signals corresponding to the cell-21 (solid curve) 
and to the spike train generated by  the R\"{o}ssler attractor fluctuations overcoming the threshold $x=7$ (circles).  {\bf b}) Autocorrelation function for the simple model telegraph signal: Eq. (A1) with $p=0.25$.  
 }
\end{figure}

\end{document}